\documentclass[12pt,a4paper]{article}
\usepackage[textwidth=15cm]{geometry}
\usepackage{authblk}
\usepackage{amsmath} 
\usepackage{bm}        
\usepackage{amssymb}   
\usepackage[small]{caption}
\usepackage{epsf,epsfig,graphicx,color,hyperref}
\usepackage[cal=cm]{mathalfa}

\newcommand{\bra}[1]{\ensuremath{\left\langle#1\right|}}
\newcommand{\ket}[1]{\ensuremath{\left|#1\right\rangle}}

\newcommand{\vevr}{\left< r \right>}
\newcommand{\vevh}{\left< \varphi \right>}

\newcommand{\bnu}{\bar{\nu}}
\newcommand{\bpsi}{\bar{\psi}}
\newcommand{\al}{\alpha}
\newcommand{\Gev}{\rm GeV}
\newcommand{\Mev}{\rm MeV}

\newcommand{\vp}{\vec{p}}

\newcommand{\vq}{\vec{q}}

\newcommand{\vqone}{\vec{q}_1}
\newcommand{\vqtwo}{\vec{q}_2}
\newcommand{\tilom}{\tilde{\omega}}
\newcommand{\twopi}{\left( 2 \pi \right)}

\begin{document}

\title{Goldstone Boson Emission From Nucleon Cooper Pairing in Neutron Stars
and Constraints on the Higgs Portal Models}

\author{Huitzu Tu\thanks{huitzu2@gate.sinica.edu.tw}}
\affil{Institute of Physics, Academia Sinica, Taipei 11529, Taiwan}

\maketitle

\begin{abstract}

When a neutron star cools to below the critical temperature for the onset of 
superfluidity, nucleon pair breaking and formation (PBF) processes become the  
dominant mechanism for neutrino emission, while the modified URCA and the nuclear
bremsstrahlung processes are suppressed.
The PBF processes in neutron stars have also been used to set upper limits on the 
properties of axions, which are comparable to those set by supernova SN 1987A.
We apply this constraint on Weinberg's Higgs portal model, in which the dark
radiation particles (the Goldstone bosons) and the dark matter candidate (a 
Majorana fermion) interact with the Standard Model (SM) fields solely through the
mixing of the SM Higgs boson and a light Higgs boson.
We compare the Goldstone boson emissivity with that of the neutrinos by considering 
several superfluid gap models for the neutron singlet-state pairing in the 
neutron star inner crust, as well as in the core region.
We find that the PBF processes in the superfluid neutron star interior can indeed 
probe Weinberg's Higgs portal model in a new parameter space region. 
Together with our previous works on the constraints from supernovae and 
gamma-ray bursts, this study demonstrates further the competitiveness and 
complementarity of astrophysics to laboratory particle physics experiments.

\end{abstract}

\maketitle

\section{Introduction}
\label{sec:intro}

After its birth, a neutron star with an initial core temperature of 
$\sim 10^{11}~{\rm K}$ cools via neutrino emission from the 
interior during the first $10^5$ years, and subsequently by photon thermal emission 
from the surface~\cite{Yakovlev:2000jp,Yakovlev:2004iq,Page:2005fq,Potekhin:2015qsa,Schmitt:2017efp,Potekhin:2017ufy,Ozel:2012wu}.
The basic neutrino production mechanisms are the modified URCA and the nuclear
bremsstrahlung processes.
The direct URCA processes are possible only if the proton fraction in the degenerate nuclear matter exceeds a certain threshold value (see e.g. Ref.~\cite{Brown:2017gxd}.)
Analytical approximations of neutrino luminosities for those three processes 
have also been derived~\cite{Ofengeim:2016rkq,Ofengeim:2017xxr}, which can be used 
to simplify the comparison of theoretical modelling with the observations of
neutron star cooling.

However, as the neutron star interior cools to below the critical temperature 
for the onset of nucleon superfluidity, theoretical calculation is much complicated.
Superfluidity and superconductivity have two important effects on neutron star
cooling via neutrino emission. 
On the one hand the modified URCA emission rate is reduced due to the 
appearance of an energy gap at the Fermi surface, which suppresses single particle
excitations of the paired nucleons. 
On the other hand, the nucleon pair breaking and formation (PBF) processes 
are switched on and become the dominant neutrino emission mechanism at this
stage~\cite{Flowers:1976ux,Voskresensky:1987hm,Yakovlev:1998wr,Leinson:2006gh,Kolomeitsev:2008mc,Steiner:2008qz,Page:2009fu,Kolomeitsev:2010hr,Kolomeitsev:2010pm,Leinson:2014cja,Leinson:2017dlo}.

It is generally assumed that neutrons undergo singlet-state Cooper pairing in the
neutron star inner crust, and triplet-state pairing in the core~\cite{Tamagaki:1970}.
Protons are expected to undergo singlet-state pairing in the core.
The NASA Chandra X-ray observatory has recorded a steady decline of the 
effective surface temperature of the supernova remnant Cassiopeia A.
This finding is interpreted as the evidence for the existence of superfluidity 
in its core~\cite{Shternin:2010qi,Elshamouty:2013nfa}.
Neutron star superfluidity may also be studied using pulsar glitches, as 
shown by Refs.~\cite{Ho:2015vza,Ho:2017ipg}.
For recent reviews on the superfluidity in neutron stars, we refer to 
Refs.~\cite{Potekhin:2015qsa,Gezerlis:2014efa,Pethick:2015jma,Haskell:2017lkl,Sedrakian:2018ydt}.

Beyond the Standard Model (SM) of particle physics,
axion emission from nucleon PBF processes in neutron stars has been studied in
Refs.~\cite{Sedrakian:2018ydt,Keller:2012yr,Leinson:2014ioa,Sedrakian:2015krq}.
Simulation of neutron star cooling by axions in addition to neutrinos performed in 
Ref.~\cite{Sedrakian:2015krq} determined an upper bound on the axion mass of
$m_a \lesssim (0.06 - 0.12)~{\rm eV}$, comparable to those set by supernovae
(SN 1987A) and white dwarfs. 
See Ref.~\cite{Sedrakian:2018ydt} for a most recent summary of this topic.
In this work we show that another interesting example to study in the context of 
neutron cooling is provided by Weinberg's Higgs portal model~\cite{Weinberg:2013kea}, 
which was proposed to account for the dark radiation in the early 
universe~\cite{Riess:2016jrr} (see, however, also Ref.~\cite{Heavens:2017hkr}.)
In this model, Weinberg considered a global $U (1)$ continuous symmetry associated
with the conservation of some quantum number, and introduced a complex scalar 
field to break it spontaneously.
The radial field of the complex scalar field acquires a vacuum expectation value
(vev), and mixes with the SM Higgs field.
The Goldstone bosons arising from the symmetry breaking would be massless and 
very weakly-interacting. 
Thus they can decouple from the early universe thermal bath at the right moment, 
and be a good dark radiation candidate.

Previously we have examined energy losses due to the resonant emission of Weinberg's 
Goldstone bosons in a post-collapse supernova core~\cite{Keung:2013mfa,Tu:2017dhl},
as well as in the initial fireballs of gamma-ray bursts~\cite{Tu:2015lwv}.
In this work we focus on the Goldstone boson resonance production when the neutron
star cools to just below the critical temperature for the onset of the 
superfluidity.
In section~\ref{sec:neutrino} we briefly summarise current attempts to study
neutrino emission from the nucleon PBF processes using the Green function 
approach in the Nambu-Gor'kov formalism. 
Section~\ref{sec:model} contains a short review on Weinberg's Higgs portal model.
In Section~\ref{sec:goldstone} we describe our method for estimating the Goldstone
boson emissivity from the nucleon PBF processes.
In section~\ref{sec:emissivity} we first determine the profiles of the neutron
superfluid gap energy and the critical temperature in the neutron star inner crust
and core region. 
Based on this information we compare the Goldstone boson and neutrino emissivities
by adopting different gap models, and at different radii inside the neutron star.
Variations of our neutron PBF bounds on Weinberg's Higgs portal model are discussed,
and the bounds are confronted with those set by laboratory experiments and 
by high-energy astrophysics.  
In section~\ref{sec:summary} we summarise this work.

\section{Neutrino Emissivities from the Nucleon Cooper Pair Breaking and Formation
Processes}
\label{sec:neutrino}

First we summarise current knowledge of the neutrino emissivities from the nucleon
Cooper pair breaking and formation (PBF) processes
\begin{equation}
   N + N \rightarrow \{N N \} + \nu + \bar{\nu}\, , \hspace{0.3cm}
   \{N N \} \rightarrow N + N + \nu + \bar{\nu}\, ,
\end{equation}
where $\{N N \}$ represents a nucleon Cooper pair, with $N = n$ (neutron) or $p$
(proton).
The low-energy effective Lagrangian for neutral weak current interactions can be 
written as
\begin{equation}
   \mathcal{L} = \frac{G_F}{2 \sqrt{2}}\, J^\mu\, l_\mu\, ,
\end{equation}
where $G_F$ is the Fermi constant, and the neutrino weak current is 
$l_\mu = \bnu \gamma_\mu\, (1 - \gamma_5) \nu$.
The nucleon weak current is 
$J_\mu = \bpsi_N \gamma_\mu\, (c_V - c_A \gamma_5) \psi_N \equiv 
J_{\mu, V} - J_{\mu, A}$, with $c_V$ and $c_A$
the vector and the axial-vector coupling, respectively.
For the reactions with neutrons, $c_V = 1$ and $c_A = g_A$, where 
$g_A \approx 1.26$ is the axial coupling constant.
In the non-relativistic limit, the nucleon weak vector and axial-vector 
currents become
\begin{eqnarray}
\label{eq:Jnonrel}
   J_{\mu, V} &\approx& \psi^\dagger_N (p_1)
   \left(1, \frac{(\vec{p}_1 + \vec{p}_2)}{2 m_N} \right) \psi_N (p_2)\, , 
   \nonumber \\
   J_{\mu, A} &\approx& \psi^\dagger_N (p_1)
   \left(\frac{\vec{\sigma} \cdot (\vec{p}_1 + \vec{p}_2)}{2 m_N}, 
   \vec{\sigma} \right) \psi_N (p_2)\, ,
\end{eqnarray}
respectively, with $\vec{\sigma} = (\sigma_1, \sigma_2, \sigma_3)$ the Pauli matrices
acting on the nucleon spinors $\psi_N$.

At finite temperature $T$, neutrino emissivity per neutrino flavour is 
calculated by means of the imaginary part of the retarded weak polarisation tensor
$\Pi_{\mu \nu}$, 
as (for derivations see e.g. Refs.~\cite{Sedrakian:2000kc,Leinson:2000un})
\begin{equation}
   Q_{\nu \bnu} = \left(\frac{G_F}{2 \sqrt{2}} \right)^2 \int 
   \frac{d^3 \vec{q}_1}{\twopi^3 2 \omega_1} 
   \frac{d^3 \vec{q}_2}{\twopi^3 2 \omega_2}\, \omega\, f_B (\omega)\, 2\,
   \text{Im}\, \Pi^{\mu \nu} (q)\, \text{Tr}\, (l_\mu l^\ast_\nu)\, .
\end{equation}
Here the four-momentum transfer is $q = q_1 + q_2 = (\omega, \vec{q})$, 
with $q_i = (\omega_i, \vec{q}_i)$, $i=1, 2$ the 
four-momenta of the two emitted neutrinos, and 
$f_B (\omega) = [\exp (\omega / T) - 1]^{-1}$ the Bose distribution function.
By inserting $\int d^4 q\, \delta^{(4)} (q - q_1 - q_2) = 1$ 
into the above equation and making use of Lenard's integral, 
the leptonic trace $\text{Tr}\, (l_\mu l^\ast_\nu)$ is integrated to be 
\begin{equation}
\label{eq:lepton}
   \int \frac{d^3 \vec{q}_1}{\twopi^3 2 \omega_1} 
   \frac{d^3 \vec{q}_2}{\twopi^3 2 \omega_2}\, \text{Tr}\, (l_\mu l^\ast_\nu)\,
   \delta^{(4)} (q - q_1 - q_2)
   = \frac{1}{48 \pi^5}\, (q_\mu q_\nu - q^2\, g_{\mu \nu})\, 
   \Theta (q^2)\, \Theta (\omega)\, .
\end{equation}

\subsection{Green functions and dressed vertices in neutron superfluid}

The retarded polarisation tensor of the $Z$ boson, $\Pi_{\mu \nu} (q)$, is to be 
calculated using the dressed weak vertices $\hat{\Gamma}^\mu$ and the nucleon 
quasi-particle propagators $\hat{G}$ (see e.g. Ref.~\cite{Schmitt:2005wg}), where
$\hat{\mathcal{O}}$ denotes operators in the Nambu-Gor'kov space.
In the Nambu-Gor'kov formalism~\cite{Nambu:1960tm}, a two-component spinor field 
operator 
\begin{equation}
   \Psi_p = \binom{c_{p\, \uparrow}}{c^\dagger_{- p\, \downarrow}}
   = \binom{\Psi_{p\, 1}}{\Psi_{p\, 2}}\, ,
\end{equation}
is introduced to study the particle-hole dynamics. 
Here $c^\dagger_{p\, \uparrow}$ and $c_{p\, \uparrow}$ are the creation and 
annihilation operators for quasi-particle of four-momentum $p$, and spin direction
$\uparrow$.
The one-particle Green's function matrix is defined as
\begin{equation}
   \hat{G}_{\alpha \beta} (t, \vp) \equiv -i \bra{0} \mathcal{T} 
   \{ \Psi_{p \alpha} (t)\, \Psi^\dagger_{p \beta} (0) \} \ket{0}\, ,
\end{equation}
for $\alpha, \beta= 1, 2$, where $\mathcal{T}$ is the time-ordering symbol.
The ground state (vacuum) $\ket{0}$ is that of the modified zero-order 
Hamiltonian~\cite{Schrieffer:1964zz}
\begin{eqnarray} 
   H^\prime_0 &=& H_0 + (H_{\rm HF} + H_\phi) - \mu N \nonumber \\
   &=& \sum_p \Psi^\dagger_p \left[ \bar{\epsilon}_p\, \hat{\sigma}_3 + 
   \phi_{p\, 1}\, \hat{\sigma}_1 + \phi_{p\, 2}\, \hat{\sigma}_2 \right] \Psi_p\, ,
\end{eqnarray}
for an average number of nucleons $N_0$. 
Here $H_{\rm HF} = \sum_{p, s} \varepsilon_{\rm HF} (p)\, n_{p, s}$ 
is the Hartree-Fock potential, where $\sum_{p, s} n_{p, s}= N$ gives the total number
of nucleons.
The term $H_\phi$ is added for describing the pairing correlations, 
and $\mu$ is the chemical potential. 
The Pauli matrices $\hat{\sigma}_i$, $i=1, 2, 3$, operate in the Nambu-Gor'kov space.
Choosing $\phi_{p\, 2} = 0$ as permitted by the rotational invariance in the 
$\hat{\sigma}$ space, $\phi_{p\, 1}$ is identified as the energy-gap parameter
$\Delta_p$.
The quasi-particle energy is then
\begin{equation}
   E_p = \sqrt{\bar{\epsilon}^2_p + \Delta^2_p}\, ,
\end{equation}
where $\bar{\epsilon}_p \equiv \epsilon_p + \varepsilon_{\rm HF} (p) - \mu$, with
$\epsilon_p = |\vp|^2 / (2 m^\ast_N)$, and $m^\ast_N$ the nucleon effective mass.

Assuming that the nuclear medium is in thermal equilibrium at temperature $T$,
one can adopt the Matsubara Green's function technique. 
All energy variables associated with fermion lines are replaced by
$p^0 \rightarrow i p_n = i \pi\, (2 n +1)\, T$, with $n=0, \pm 1, \pm 2, ...$. 
The boson energy variables are similarly replaced by
$q^0 \rightarrow i \omega_m = i \pi\, 2 m\, T$, with $m=0, \pm 1, \pm 2, ...$.

Consider the singlet-state ($^{2 S + 1} L_J =$ $^1 S_0$) pairing of neutrons first.
In the spectroscopic notation specifying the nucleon-nucleon scattering,  
$\vec{S}$ is the total spin, $\vec{L}$ the total orbital angular momentum, with
$L = 0, 1, 2, 3, ... \Rightarrow S, P, D, F, ...$, and $\vec{J} = \vec{L} + \vec{S}$
the total angular momentum.  
In the frequency-momentum space, the Green's function for nucleons is 
\begin{equation}
\label{eq:green}
   \hat{G}_{\uparrow} (p_n, \vp) = \left( 
   \begin{array}{cc}
   G_\uparrow (p_n, \vp) & F_\uparrow (p_n, \vp) \\
   F^\dagger_\uparrow (-p_n, \vp) & - G_\downarrow (-p_n, -\vp) \\
   \end{array} \right)\, . 
\end{equation}
The diagonal elements 
\begin{equation}
\label{eq:green_normal}
    G_\uparrow (p_n, \vp) = \frac{- i p_n - \bar{\epsilon}_p}{p^2_n + E^2_p}\, ,
    \hspace{0.6cm}
    G_\downarrow (-p_n, -\vp) = G^\dagger_\uparrow (p_n, \vp) =
    \frac{i p_n - \bar{\epsilon}_p}{p^2_n + E^2_p}\, ,
\end{equation}
are the normal propagators for the particle and the hole, respectively.
Their relation is determined by the space and spin inversion symmetry properties
of the system, and in the above equation is that in the singlet-state pairing case.
The off-diagonal elements 
\begin{equation}
\label{eq:green_anomalous}
    F_\uparrow (p_n, \vp) = \frac{\Delta_p}{p^2_n + E^2_p}\, ,
    \hspace{0.6cm}
    F^\dagger_\uparrow (-p_n, \vp) = F_\uparrow (p_n, \vp)\, ,
\end{equation}
are termed the anomalous propagators, which describe the transition of a particle
into a hole and a condensate pair (and vice versa) in the $^1 S_0$ pairing case.
The $Z$ boson polarisation tensor or self-energy is 
then~\cite{Schrieffer:1964zz,Leinson:2008ba}  
\begin{eqnarray}
   \Pi_{\mu \nu} (\omega_m, \vq) &=& T \sum_{p_n} \int \frac{d^3 \vp }{\twopi^3}\, 
   \text{Tr}\, [\hat{\gamma}_\mu (p_n, \vp; p_n + \omega_m, \vp + \vq) \nonumber \\
   && \times\, \hat{G} (p_n + \omega_m, \vp + \vq)\, 
   \hat{\Gamma}_\nu (p_n + \omega_m, \vp + \vq; p_n ,\vp)\, \hat{G} (p_n ,\vp)]\, ,
\end{eqnarray}
where $p$ and $p+q$ are the fermion four-momenta in the loop.
Analytic continuation of the polarisation tensor with retarded 
boundary condition, $i \omega_m \rightarrow \omega + i \eta$, gives rise to 
a delta function of $\delta (\omega^2 - 4 E^2_p)$ for the imaginary part of the 
retarded weak polarisation tensor, ${\rm Im} \Pi_{\mu \nu} (q)$.

In the Nambu-Gor'kov formalism, the bare vector vertex can be written as
\begin{eqnarray} 
\label{eq:vertex_vec}
   \hat{\gamma}_\mu (p + q, p) &=& \left(
   \begin{array}{cc}
   \gamma_\mu (p + q, p) & 0 \\
   0 & - \gamma^\dagger_\mu (p + q, p) \\
   \end{array} \right) \nonumber \\  
   &=& \begin{cases} 
   \hat{\sigma}_3 & \hspace{0.5cm} \text{if} \hspace{0.2cm} \mu = 0\, , \\
   \frac{1}{m^\ast_N}\, \left(\vp + \frac{\vq}{2} \right) 
   \hat{\mathcal{I}} & \hspace{0.5cm} 
   \text{if} \hspace{0.2cm} \mu = i = 1, 2 ,3\, ,
   \end{cases}
\end{eqnarray} 
where $\hat{\mathcal{I}}$ is the identity matrix in the Nambu-Gor'kov space.
The dressed vector vertex is determined by the Bethe-Salpeter equation
\begin{equation}
\label{eq:bethe_salpeter}
   \hat{\Gamma}_\mu (p + q, p) = \hat{\gamma}_\mu (p + q, p) +
   i \int \frac{d^4 k}{\twopi^4}\, \hat{\sigma}_3\, \hat{G} (k + q)\, \hat{\Gamma}_\mu
   (k + q, k)\, \hat{G} (k)\, \hat{\sigma}_3 \mathcal{V} (p - k)\, , 
\end{equation}
where $\mathcal{V}$ is the attractive two-body potential between the nucleons.
Here the Green's function for the interacting system is approximated by using the 
Dyson equation
\begin{equation}
   \hat{G}^{-1} (p) = p^0\, \hat{\mathcal{I}} - \epsilon_p\, \hat{\sigma}_3 - 
   \Sigma (p)\, ,
\end{equation}
where the self-energy $\Sigma (p)$ is a functional of the Green's function,
$\Sigma = \Sigma [G]$.
Note that the relation between the dressed vertex function and the Green's function
is governed by the generalised Ward identity~\cite{Nambu:1960tm,Schrieffer:1964zz}
\begin{equation}
\label{eq:ward}
   q^\mu \hat{\Gamma}_\mu (p+q, p) = \hat{G}^{-1} (p+q) \hat{\sigma}_3 - 
   \hat{\sigma}_3 \hat{G}^{-1} (p)\, ,
\end{equation}
so that gauge invariance is ensured.

\subsection{Neutrino emissivity: vector and axial-vector current contributions 
in different pairing states}

The mixed terms between the vector and axial-vector terms are antisymmetric, 
therefore they vanish after the contraction with the symmetric leptonic current 
tensor $(q_\mu q_\nu - q^2 g_{\mu \nu})$ in Eq.~(\ref{eq:lepton}).
It has been pointed out that at densities higher than the nuclear saturation
density, triplet-state ($P$-wave state $^3 P_2$ mixed with $F$-wave state 
$^3 F_2$) pairing is more favourable~\cite{Takatsuka:1992ga}.
Following Refs.~\cite{Levenfish:1994a,Levenfish:1994b}, one usually considers three 
types of BCS superfluidity:
singlet-state $^1 S_0$, triplet-state $^3 P_2$ with $m_J = 0$, and 
triplet-state $^3P_2$ with $|m_J| = 2$.
Vector-current contribution in the case of the $^1 S_0$ pairing for 
${\mathcal N}_\nu$ neutrino species
is found by various approaches to 
be~\cite{Leinson:2006gh,Kolomeitsev:2010hr,Sedrakian:2012ha} 
\begin{equation}
\label{eq:QnusV}
   Q^{(s)}_{\nu \bnu, \, V} \approx 0.013\,
   \frac{G^2_F\, m^\ast_N\, p_{\rm F}}{3 \pi^5}\, c^2_V\, {\mathcal N}_\nu\,
   v^4_F\, \Delta_p^2\, 
   \int^\infty_{\Delta_p} \frac{d E_p\, E_p^5}{\sqrt{E_p^2 - \Delta_p^2}} 
   \frac{1}{(e^{E_p/T} + 1)^2}\, .
\end{equation}
The neutron Fermi momentum is $p_F = (3 \pi^2 n_n)^{1/3}$, with $n_n$ the 
neutron number density, and the neutron Fermi velocity is defined by 
$v_F = p_F / m^\ast_N$.
In Refs.~\cite{Flowers:1976ux,Yakovlev:1998wr} first estimates for 
$Q^{(s)}_{\nu \bar{\nu}, V}$ were made by using the non-relativistic approximation
$J_{\mu, V} \approx \psi^\dagger_N \psi_N$ in Eq.~(\ref{eq:Jnonrel}).
In the calculations therein only the temporal component of the bare vector vertices 
was considered, in which case the generalised Ward identity, Eq.~(\ref{eq:ward}), 
could not be satisfied.
The result was an overestimation of the neutrino emissivity.
Up to now different approaches agree that the early 
estimates~\cite{Flowers:1976ux,Yakovlev:1998wr} are reduced by a factor of 
about $0.05\, v^4_F$.

Axial-vector current contribution in the case of singlet-state pairing was first
considered in Ref.~\cite{Kaminker:1999ez}.
Other estimates~\cite{Kolomeitsev:2010hr,Leinson:2009mq,Sedrakian:2012mv} 
all agree with it within $25\%$, and can be approximated by
\begin{equation}
\label{eq:QnusA}
   Q^{(s)}_{\nu \bnu, \, A} \approx 0.23\,
   \frac{G^2_F\, m^\ast_N\, p_{\rm F}}{\pi^5}\, g^2_A\, {\mathcal N}_\nu\,
   v^2_F\, \Delta_p^2\, 
   \int^\infty_{\Delta_p} \frac{d E_p\, E_p^5}{\sqrt{E_p^2 - \Delta_p^2}} 
   \frac{1}{(e^{E_p/T} + 1)^2}\, ,
\end{equation}
assuming an effective nucleon mass $m^\ast_N = 0.7\, m_N$.
Being suppressed by a factor $ \propto v^2_F$ relative to $v^4_F$, 
the axial-vector current contribution thus dominates over the vector one.

Neutrino emissivity due to the triplet-state $(^3 P_2)$ Cooper pairing of neutrons 
was first calculated in Ref.~\cite{Yakovlev:1998wr}, and then more recently in 
Refs.~\cite{Leinson:2017dlo,Leinson:2009nu}.
It is generally assumed that the vector-current contribution is negligible, and
the axial-vector current contribution is that obtained in Ref.~\cite{Yakovlev:1998wr}
suppressed by a factor of $1/4$ (for $|m_J| = 0$) or $1/8$ (for $|m_J| = 2$).
In the case of triplet-state pairing, the gap energy $\Delta_{\vp}$ depends on the 
direction of the quasi-particle momentum $\vp$.
The expression for the $^3 P_2$ pairing with $|m_J| = 0$ 
is~\cite{Leinson:2006gh,Leinson:2010pk,Leinson:2011jr,Leinson:2016dat}
(see also Refs.~\cite{Page:2009fu,Sedrakian:2015krq})
\begin{equation}
   Q^{(t\, , |m_J|=0)}_{\nu \bnu, \, A} \approx 
   \frac{2 G^2_F\, m^\ast_N\, p_{\rm F}}{15\, \pi^5}\, c^2_A\, 
   {\mathcal N}_\nu\, \frac{1}{4 \pi} \int d \Omega\,
   \Delta^2_{{\vp}}\, \int^\infty_{\Delta_{\vp}} \frac{d Ep\, E_p^5}
   {\sqrt{E_p^2 - \Delta^2_{\vp}}} \frac{1}{(e^{E_p/T} + 1)^2}\, .
\end{equation}
For the case of $|m_J|=2$, the neutrino emissivity is 
$Q^{(t\, , |m_J|=2)}_{\nu \bar{\nu}, \, A} = 
0.5\, Q^{(t\, , |m_J|=0)}_{\nu \bar{\nu}, \, A}$.
In the non-relativistic limit the axial-vector current represents the nucleon spin
density (cf. Eq.~(\ref{eq:Jnonrel}).)
Since the nucleon spins can fluctuate freely in the pairing state where the total 
spin $S \neq 0$, the axial-vector current contributions in the triplet-state pairing
are not suppressed by the Fermi velocity $v_F$.
Thus they are regarded as the dominant neutron star cooling channels when the
core temperature drops below the critical temperature for the onset
of triplet-state pairing of neutrons, $T_{cnt}$.

\section{Weinberg's Higgs Portal Model}
\label{sec:model}

In this section we briefly summarise Weinberg's model~\cite{Weinberg:2013kea} 
following the convention of Refs.~\cite{Keung:2013mfa,Cheung:2013oya}.
Consider the simplest possible broken continuous symmetry, a global $U (1)$ symmetry
associated with the conservation of some quantum number $W$.
A single complex scalar field $S (x)$ is introduced for breaking this symmetry
spontaneously. 
With this field added to the Standard Model (SM), the Lagrangian is 
\begin{equation}
\label{eq:Lagrangian1}
   {\mathcal L} = \left(\partial_\mu S^\dagger \right) \left(\partial^\mu S \right)
   + \mu^2 S^\dagger S - \lambda (S^\dagger S)^2 - g_H\, (S^\dagger S) 
   (\Phi^\dagger \Phi) + {\mathcal L}_{\rm SM}\, ,
\end{equation}
where $\Phi$ is the SM Higgs doublet, $\mu^2$, $g_H$, and $\lambda$ are real constants,
and $\mathcal{L}_{\rm SM}$ is the usual SM Lagrangian.
One separates a massless Goldstone boson field $\alpha (x)$ and a massive radial field
$r (x)$ in $S (x)$ by defining
\begin{equation}
   S (x) = \frac{1}{\sqrt{2}} \left(\vevr + r (x) \right)\, e^{2 i \al (x)}\, ,
\end{equation}
where the fields $\alpha (x)$ and $r (x)$ are real.
In the unitary gauge one sets $\Phi^{\rm T} = \left(0, \vevh + \varphi (x) \right) 
/\sqrt{2}$, where $\varphi (x)$ is the physical Higgs field.
The Lagrangian in Eq.~(\ref{eq:Lagrangian1}) thus becomes
\begin{eqnarray}
   \mathcal{L} &=& \frac{1}{2} \left(\partial_\mu r \right) 
   \left(\partial^\mu r \right)
   + \frac{1}{2} \frac{\left(\vevr + r \right)^2}{\vevr^2} 
   \left(\partial_\mu \al \right) \left(\partial^\mu \al \right) +
   \frac{\mu^2}{2} \left(\vevr + r \right)^2 \nonumber \\
   && - \frac{\lambda}{4} \left(\vevr + r \right)^4 - \frac{g_H}{4}
   \left(\vevr + r \right)^2 \left(\vevh + \varphi \right)^2 + \mathcal{L}_{\rm SM}\, ,
\end{eqnarray}
where the replacement $\al (x) \rightarrow \al (x) / \left( 2 \vevr \right)$ was made 
in order to achieve a canonical kinetic term for the $\al (x)$ field.
The two fields $\varphi$ and $r$ mix due to the 
$g_H\, (S^\dagger S) (\Phi^\dagger \Phi)$ term, with their mixing angle given by
\begin{equation}
   \theta_H \approx \frac{g_H \vevh \vevr}{m^2_H - m^2_h}\, ,
\end{equation}
where $m_H$ and $m_h$ are the masses of the two resulting physical Higgs bosons 
$H$ and $h$, respectively.
The heavier one is identified with the SM Higgs boson with mass $m_H = 125~\Gev$, 
and vacuum expectation value (vev) of $\vevh = 246~\Gev$.
The lighter one is assumed to have a mass in the range of MeV to hundreds of MeV. 
The vev of the radial field $r$ must satisfy the perturbativity 
condition on the quartic self-coupling of the $S$ field,
\begin{equation}
\label{eq:vevrperturbativitybound}
   \lambda = \frac{m^2_h}{\vevr^2} \leq 4 \pi\, .
\end{equation}

In this model, the interaction of the Goldstone bosons with the SM fields arises 
entirely through the SM Higgs boson in the mixing of the $\varphi$ and $r$ fields. 
The Lagrangian for the interaction of the Goldstone bosons with the SM fields contains
the Yukawa terms
\begin{equation}
   \frac{m_f}{\vevh} H \bar{f} f \cos \theta_H - 
   \frac{m_f}{\vevh} h \bar{f} f \sin \theta_H\, ,
\end{equation}
and the terms which couple the Goldstone boson pair with the SM Higgs boson 
and the light Higgs boson
\begin{equation}
   \frac{1}{\vevr} H\, (\partial \al)^2 \sin \theta_H + 
   \frac{1}{\vevr} h\, (\partial \al)^2 \cos \theta_H\, .
\end{equation}

This model can also be extended to include a dark matter candidate by adding one
Dirac field
\begin{equation}
   \mathcal{L}_\psi = i \bar{\psi} \gamma \cdot \partial \psi - 
   m_\psi \bar{\psi} \psi - \frac{f_\chi}{\sqrt{2}} \bar{\psi}^c \psi S^\dagger - 
   \frac{f^\ast_\chi}{\sqrt{2}} \bar{\psi} \psi^c S\, , 
\end{equation}
and assigning a charge $U (1)_W = 1$ for it.
After the radial field $r$ acquires a vev, diagonalising the $\psi$ mass matrix
generates two mass eigenvalues $m_{\pm} = m_\psi \pm f_\chi\, \vevr$ for the 
two mass eigenstates $\psi_{\pm}$, which are Majorana fermions.
The Lagrangian contains the term
\begin{equation}
   - \frac{f_\chi}{2}\, (\cos \theta_H\, h + \sin \theta_H\, H)\, 
   \left(\bar{\psi}_+ \psi_+ - \bar{\psi}_- \psi_- \right)\, ,
\end{equation} 
for their interactions with the SM and the light Higgs bosons.
The lighter fermion is stable due to unbroken reflection symmetry, thus can play
the role of the dark matter, with mass $m_- \equiv M_\chi$ in the GeV to TeV range.
Its relic density has been calculated in Ref.~\cite{Anchordoqui:2013pta}.

Model parameters in the minimal set-up are the Higgs portal coupling $g_H$, 
mass of the light Higgs boson $m_h$, and its vev $\vevr$. 
In the extended version the WIMP dark matter mass $M_\chi$, and its 
coupling to the Higgs bosons $f_\chi$ are included.
From the SM Higgs invisible decay width, a collider bound on the Higgs portal 
coupling has been derived in Ref.~\cite{Cheung:2013oya}
\begin{equation}
\label{eq:gcolliderbound}
   g_H < 0.011\, .
\end{equation}
Laboratory experimental limits on meson invisible decay widths have also been 
turned into constraints on the $\varphi$--$r$ mixing angle $\theta_H$ in 
Ref.~\cite{Anchordoqui:2013bfa}.
We have scrutinised the production and propagation of Weinberg's Goldstone bosons
in the post-collapse supernova core~\cite{Keung:2013mfa,Tu:2017dhl}, 
as well as in the initial fireballs of gamma-ray bursts~\cite{Tu:2015lwv}. 
We found that the supernova bound based on the SN 1987A neutrino burst observations
surpasses those set by laboratory experiments for all light Higgs boson masses
$m_h \lesssim 500~\Mev$. 
Using generic values for the GRB initial fireball energy, temperature 
$T_0 \lesssim 2 \cdot 10^{11}~{\rm K}$, radius, 
expansion rate, and baryon number density, we found that the GRB bounds are 
competitive to current constraints set by $B$ meson and radiative Upsilon 
invisible decays in the parameter range $m_h / T_0 \lesssim 10$--$15$.

In Weinberg's Higgs portal model including the dark matter candidate, exclusion 
limits on the WIMP-nucleon elastic cross section set by the null results of the
direct search experiments have been found to put very strong bounds on the 
$\varphi$--$r$ mixing angle~\cite{Anchordoqui:2013bfa}.

\section{Goldstone Boson Emissivity from the Nucleon Cooper Pair Breaking and 
Formation Processes}
\label{sec:goldstone}

The effective Lagrangian for the interactions of the Goldstone boson with the nucleons 
is given by
\begin{equation}
   \mathcal{L} = \left(\frac{f_N g_H\, m_N}{m^2_H m^2_h} \right)\, \bpsi_N \psi_N\,
   \partial_\mu \al \partial^\mu \al\, . 
\end{equation}
The Higgs effective coupling to nucleons, $f_N m_N /\vevh \equiv g_{N N H}$, has been
calculated for the purpose of investigating the sensitivities of the dark matter 
direct detection experiments.
So far various estimates agree that $f_N \simeq 0.3$
(see e.g. Ref.~\cite{Tu:2017dhl} for a brief summary.)

For the Goldstone boson emissivity from the neutron PBF processes
\begin{equation}
   N + N \rightarrow \{N N \} + \al + \al\, \hspace{0.4cm}
   \{N N \} \rightarrow N + N + \al + \al\, ,
\end{equation}
we calculate by means of the imaginary part of the Higgs boson self-energy $\Pi$, as
\begin{equation}
   Q_{\al \al} = \frac{1}{2!} \left(\frac{f_N g_H\, m_N}{m^2_H} \right)^2 \int 
   \frac{d^3 \vec{q}_1}{\twopi^3 2 \omega_1} 
   \frac{d^3 \vec{q}_2}{\twopi^3 2 \omega_2}\, \omega f_B (\omega)\, 2\,
   \text{Im}\, \Pi (q)\, 
   \frac{(q_1 \cdot q_2)^2}{(q^2 - m^2_h)^2 + m^2_h \Gamma^2_h}\, ,
\end{equation}
in analogy to that of the neutrinos.
Here $q = q_1 + q_2 = (\omega, \vec{q})$, with $q_i = (\omega_i, \vec{q}_i)$, 
$i=1, 2$ the four-momenta of the two emitted Goldstone bosons, and 
$f_B (\omega) = [\exp (\omega / T) - 1]^{-1}$ the Bose distribution function.
A symmetry factor of $1 / 2!$ is included to take into account the two identical 
particles in the final state.
Here we use a Breit-Wigner form for the propagator of the intermediate light Higgs 
boson $h$. 
It decays dominantly to a pair of Goldstone bosons, with the decay width given by
\begin{equation}
\label{eq:Gammah}
   \Gamma_h = \frac{1}{32 \pi} \frac{m^3_h}{\vevr^2}\, .
\end{equation}
When kinematically allowed, there is also a probability for $h$ decaying into a 
pair of SM fermions, as well as a pair of pions.

\subsection{Resonance effect in Goldstone boson production}

We evaluate the integral over the Goldstone boson momenta first:
\begin{equation}
  \int \frac{d^3 \vqone}{\twopi^3 2 \omega_1} 
  \frac{d^3 \vqtwo}{\twopi^3 2 \omega_2} 
  \frac{(q_1 \cdot q_2)^2}{(q^2 - m^2_h)^2 + m^2_h \Gamma^2_h}\,
  = \frac{1}{2 {\twopi}^4} 
  \int^\infty_0 d \omega\, \omega^3\, I_\al (\omega, m_h, \vevr)\, .
\end{equation}
The dimensionless integral is defined by
\begin{equation}
\label{eq:Ial}
   I_\al (\omega, m_h, \vevr) \equiv \int^1_0 d \tilom \int^{+1}_{-1} 
   \frac{d \cos \theta_\al\, \tilom^3\, (1 - \tilom)^3\, (1- \cos \theta_\al)^2}
   {[2 \tilom\, (1 - \tilom)\, (1- \cos \theta_\al) - \frac{m^2_h}{\omega^2}]^2 + 
   \frac{m^2_h\, \Gamma^2_h}{\omega^4}}\, ,
\end{equation}
with $\tilom \equiv \omega_1 / \omega$, and $\theta_\al$ is the angle between the two 
emitted Goldstone bosons.
In the resonance region, we make use of the limit of the Poisson kernel and obtain
\begin{equation}
\label{eq:PkI1}
   I^{\rm Pk}_\al (\omega, m_h, \vevr) \approx 
   \frac{\pi}{8} \frac{m^3_h}{\Gamma_h\, \omega^2} 
   \propto \frac{\vevr^2}{\omega^2}\, .
\end{equation}
Since this approximation is valid when $m^2_h / \omega^2 \approx 2 
\tilom \left(1 - \tilom \right)$, where the latter $\leq 1$, it is only applicable
for $\omega \gtrsim \sqrt{2} m_h$ and $\Gamma_h \ll \omega$.
On the other hand, in the large $m_h$ limit, the dimensionless integral in
Eq.~(\ref{eq:Ial}) is simply 
\begin{equation}
\label{eq:largemhI1}
   I^{m_h \gg \omega}_\al \approx \frac{\omega^4}{35\, m^4_h}\, .
\end{equation}
As we will show in the next section, useful constraints on Weinberg's Higgs portal
model can only be obtained from the nucleon PBF processes when the Goldstone boson
emission is resonantly enhanced.
The typical energy of the Goldstone boson pairs emitted from the 
nucleon PBF processes is $\omega = 2 E_p \sim \mathcal{O} (1) \Delta_p$, 
therefore we expect that the production of a real light Higgs boson of mass 
$m_h \lesssim 15~\Mev$ should dominate.

\subsection{Suppression of scalar particle production in neutron superfluid}

Assuming that the nuclear medium is in thermal equilibrium at temperature $T$,
the Higgs boson self-energy can be calculated by adopting the Matsubara Green's
function technique,
\begin{equation}
   \Pi (\omega_m, \vq) = T \sum_{p_n} \int \frac{d^3 \vp }{\twopi^3}\, 
   \text{Tr}\, [\hat{\gamma}\, \hat{G} (p_n + \omega_m, \vp + \vq)\, 
   \hat{\Gamma} (p_n + \omega_m, \vp + \vq; p_n ,\vp)\, 
   \hat{G} (p_n ,\vp)]\, . \\
\end{equation}
Here all energy variables associated with fermion lines are replaced by
$p^0 \rightarrow i p_n = i \pi\, (2 n +1)\, T$, with $n=0, \pm 1, \pm 2, ...$. 
The boson energy variables are similarly replaced by
$q^0 \rightarrow i \omega_m = i \pi\, 2 m\, T$, with $m=0, \pm 1, \pm 2, ...$.
The bare scalar vertex is $\hat{\gamma} = \hat{\sigma}_3$ in the Nambu-Gor'kov space
(cf. Eq.~(\ref{eq:vertex_vec}).)
The quasi-particle Green's function $\hat{G}$ is given in Eqs.~(\ref{eq:green}), 
(\ref{eq:green_normal}), and (\ref{eq:green_anomalous}) in the case of 
singlet-state pairing.
Analytic continuation of the polarisation tensor with retarded 
boundary condition, $i \omega_m \rightarrow \omega + i \eta$, gives rise to a delta
function of $\delta (\omega^2 - 4 E^2_p)$ for the imaginary part of the 
retarded scalar polarisation tensor, ${\rm Im} \Pi (q)$.

The dressed scalar vertex $\hat{\Gamma}$ in dense nuclear medium is determined 
by the Bethe-Salpeter equation similarly to Eq.~(\ref{eq:bethe_salpeter}) 
for $\mu = 0$.
However, unlike the vector vertex, the scalar vertex is not protected by the 
generalised Ward identity given in Eq.~(\ref{eq:ward}).
Therefore there is a finite shift, as pointed out in e.g. Ref.~\cite{Serot:1994xh}.
In Ref.~\cite{Saito:2003js} the form factors for the strong interaction vertices of 
the scalar and the vector meson with the nucleons in dense nuclear matter 
(in the normal state) 
\begin{equation}
   F_a (Q^2, \rho_B) = R_a (Q^2, \rho_B)\, F^{\rm vac}_a (Q^2)\, ,
\end{equation}
are calculated in dependence of the momentum transfer $Q^2$ and nuclear density
$\rho_B$.
Here $a = s, v$ for the scalar and the vector meson, respectively, and
$F^{\rm vac}_a (Q^2)$ are the corresponding empirically determined form factors 
in vacuum.
For the scalar form factor $F_s (Q^2, \rho_B)$, the reduction factor 
$R_s (Q^2, \rho_B)$ contains 
a shift which does not vanish for momentum transfer $Q^2 \rightarrow 0$.
It is found that $R_s \approx R_v$ at normal nuclear matter density 
$\rho_B \sim \rho_0 = 0.17~{\rm fm}^{-1}$ and at momentum transfer 
$Q^2 \lesssim 1 ~\Gev^2$.

Ref.~\cite{Kolomeitsev:2008mc,Leinson:2008ba,Sedrakian:2012ha} 
provides the expressions for the temporal $(\propto \hat{\Gamma}^0)$ and spatial 
$(\propto \hat{\Gamma}^i, i=1, 2, 3)$ components of the neutrino vector-current 
emissivity from the nucleon PBF processes (cf. Eq.~(\ref{eq:QnusV}))
\begin{equation}
    Q^{(s)}_{\nu \bar{\nu}, V} = \left(\frac{G_F}{2 \sqrt{2}} \right)^2
    c^2_V \int^\infty_0 d \omega\, \omega f_B (\omega) \int^\omega_0
    \frac{d |\vq|\, |\vq|^2}{6 \pi^4} K_V (|\vq|, \omega)\, ,
\end{equation}
where $K_V (|\vq|, \omega) = (|\vq|^2 / \omega^2 - 1) (K_{V, 0} (|\vq|, \omega)
- K_{V, 1} (|\vq|, \omega))$. 
Both the temporal and the spatial components are of the order $v^4_F$,
\begin{eqnarray}
   K_{V, 0} &\propto& \frac{4}{45} \frac{|\vq|^4}{\omega^4} v^4_F\, , \nonumber \\
   K_{V, 1} &\propto& \frac{2}{9} \frac{|\vq|^2}{\omega^2} v^4_F\, .
\end{eqnarray} 
On the other hand, by keeping only the first term containing the temporal component
in the full $K_V (|\vq|, \omega)$ expansion, 
which is $(|\vq|^2 / \omega^2) K_{V, 0} (|\vq|, \omega)$, 
one obtains exactly the same value for $Q^{(s)}_{\nu \bar{\nu}, V}$.
In the non-relativistic limit, the temporal component of the vector-current
interaction can represent scalar interaction.
There should be finite shifts in the temporal and the spatial components of the 
vector-current contribution.
However, the Ward identity guarantees that both shifts cancel each other when
the temporal and the spatial components are added together.

Since the shift in the case we are considering is unknown, we make the assumption
that in the superfluid dense nuclear medium, the scalar vertex is modified by the 
same amount as the temporal component of the vector vertex.
Goldstone boson emissivity in the singlet-state $(^1 S_0)$ pairing is then
\begin{equation}
\label{eq:Q_als}
   Q^{(s)}_{\al \al} \approx F_s
   \left(\frac{f_N g_H m_N}{m^2_H} \right)^2 \frac{m^\ast_N\, p_{\rm F}}{4 \pi^5}\, 
   \Delta^2_p\, 
   \int^\infty_{\Delta_p} \frac{d E_p\, E^3_p}{\sqrt{E^2_p - \Delta^2_p}} 
   \frac{I_\al (\omega= 2E_p, m_h, \vevr)}{(e^{E_p / T} + 1)^2}\, ,
\end{equation}
with the suppression factor $F_s \sim 0.05\, v^4_F$ from Eq.~(\ref{eq:QnusV}).

In the case of triplet-state pairing, the Goldstone boson emissivity is also 
suppressed, similar to the vector-current contribution for the neutrino emissivity.

\section{Constraints on Weinberg's Higgs Portal Model}
\label{sec:emissivity}

In this section we evaluate the Goldstone boson emissivity from the nucleon 
PBF processes, and compare with those of the neutrinos. 
It is necessary to know in which regions of the neutron star interior and at what
temperature those processes are most efficient.

\subsection{Neutron superfluid gap energies and critical temperatures}

The gap energy $\Delta_p$ depends on the temperature, the quasi-particle momentum, 
as well as the type of superfluidity. 
It is well understood at densities less than one tenth of the nuclear
saturation density, while significant uncertainties at higher densities arise 
due to complicated nature of nucleon-nucleon interactions, and difficulties in solving 
many-body problem.
Parametrisations for the critical temperatures $T_c$ and/or the gap energies 
$\Delta_p$ at the neutron Fermi momentum $p_F$ can be found in 
Refs.~\cite{Kaminker:2001ag,Kaminker:2001eu,Andersson:2004aa,Ho:2014pta}, 
which have the general form 
\begin{equation}
   \Delta (p_F) = \Delta_0 \frac{(p_F - k_0)^2}{(p_F - k_0)^2 + k_1}
   \frac{(p_F - k_2)^2}{(p_F - k_2)^2 + k_3}\, .
\end{equation}
The fit parameters $\Delta_1$, $k_0$, $k_1$, $k_2$, and 
$k_3$ are determined for various superfluid gap models in the literature.
For some representative nuclear equation of states (see e.g. 
Ref.~\cite{Oertel:2016bki} for a recent review), the phenomenological 
superfluid gap models are tested against the cooling behaviour of the 
supernova remnant Cassiopeia A in Ref.~\cite{Ho:2014pta}.

We use the C++ package LORENE~\cite{Lorene} with various equations of state (EoS)
supplements to obtain the density profile of a non-rotating neutron star 
of gravitational mass $M_g = 1.41 M_\odot$.
For the DD2 equation of state~\cite{Hempel:2009mc,Typel:2009sy,Fischer:2013eka},
the circumferential equatorial radius of this neutron star is determined to be
$R_{\rm circ} = 13.24~{\rm km}$.
Ref.~\cite{Steiner:2007rr} defines the inner crust of a cold neutron star to be
the region between the density where neutrons drip out of nuclei 
($\rho \simeq 4 \cdot 10^{11}~{\rm g / cm^3}$), and the density for the transition 
to homogeneous nucleonic matter at about half of the nuclear saturation density
($\rho \simeq 0.5\, \rho_0 \simeq 1.5 \cdot 10^{14}~{\rm g / cm^3}$).
Following this definition, the inner crust of this neutron star configuration is 
the region at the radius $9.75~{\rm km} \lesssim r \lesssim 10.8~{\rm km}$.  
The proton fraction decreases from $Y_p \simeq 0.036$ at $r = 9.7~{\rm km}$ to 
$Y_p \simeq 0.016$ at $r \simeq 10.1~{\rm km}$, and then increases to
to $Y_p \simeq 0.32$ at $r = 10.7~{\rm km}$.

\begin{center}
\begin{figure}[t!]
\includegraphics[width=0.6\textwidth,angle=-90]{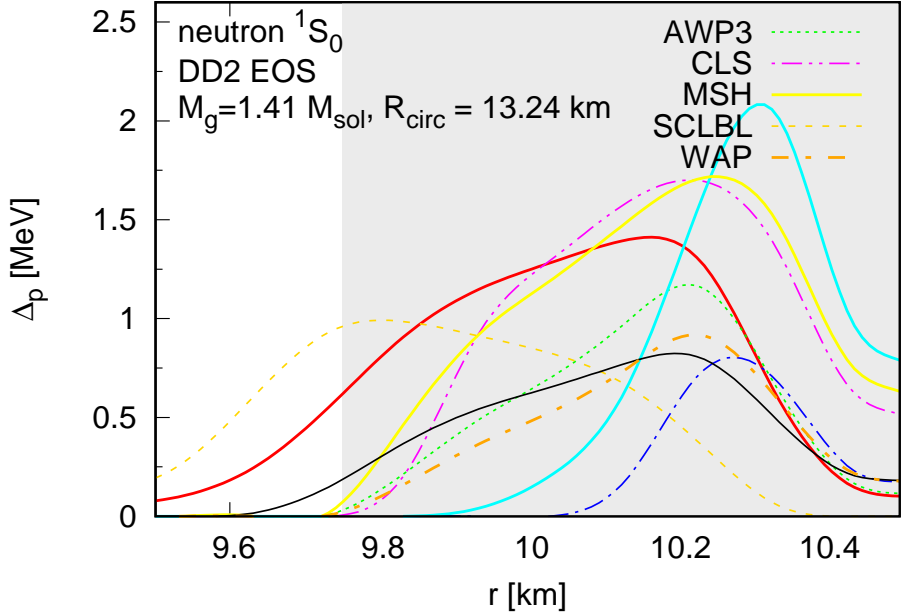}
\caption{Profile of the singlet-state ($^1S_0$) pairing gap energy 
$\Delta_p (p_F)$ inside a neutron star of gravitational mass 
$M_g = 1.41\, M_\odot$, for the nine gap models listed in Ref.~\cite{Ho:2014pta}.
The neutron star energy density profile $\rho (r)$ is calculated using the C++
package LORENE supplemented with the DD2 equation of state, then related to the 
local Fermi momentum $p_F (r)$.
Most neutron singlet-state pairing gap models are confined to the inner crust, 
which for the DD2 equation of state is the region 
$9.75 \lesssim r \lesssim 10.8~{\rm km}$ (grey shaded).
}
\label{fig:Egap_r}
\end{figure}
\end{center}

Fig.~\ref{fig:Egap_r} shows the profile of the neutron singlet-state $(^1 S_0)$ 
pairing gap energy $\Delta_p (r)$ for the nine superfluidity gap models summarised  
in Ref.~\cite{Ho:2014pta}. 
For a collection of more recent gap models, see e.g. Ref.~\cite{Sedrakian:2018ydt}.
The local Fermi momentum is related to the nuclear density $\rho (r)$ by 
$p_F (r) = (3 \pi^2 Y_n\, \rho (r) / m^\ast_N)^{1/3}$, where $Y_n = 1 - Y_p$ is 
the neutron fraction.
Most of the neutron singlet-state pairing gap models are confined to the inner 
crust, with three models extending into the core region.
As proposed by Ref.~\cite{Ho:2015vza}, these models are useful for explaining the
observed pulsar glitches.

The dependence of the gap energy on the temperature $T$ and superfluidity type is
parametrised in Ref.~\cite{Yakovlev:1998wr} as
\begin{equation}
   \Delta^2 (T, \theta) = \Delta^2_1 (T) F (\theta)\, ,
\end{equation}
where $\theta$ is the angle between the quasi-particle momentum $\vp$ 
and the quantisation axis in different pairing states.
The dependences are $F (\theta) = 1$ for the $^1 S_0$ pairing, 
$(1 + 3 \cos^2 \theta)$ for the $^3 P_2$ pairing with $m_J = 0$, and
$\sin^2 \theta$ for the $^3 P_2$ pairing with $|m_J| = 2$.
Ref.~\cite{Sedrakian:2018ydt} provides a detailed discussion on the temperature
dependence of the gap function. 
For the singlet-state pairing, the high-temperature asymptotic is
\begin{equation}
   \Delta_1 (T) \sim 3.06\, [T_c\, (T_c - T)]^{1/2}\, , \hspace{0.5cm}
   {\rm for} \hspace{0.1cm} T \rightarrow T_c\, ,
\end{equation}
somewhat larger than the earlier analytical fit given in Ref.~\cite{Yakovlev:1998wr}.

Measurement of the rapid cooling of Cassiopeia A provides the first constraints on the
critical temperature for the onset of superfluidity of core neutrons and 
protons~\cite{Shternin:2010qi,Page:2010aw}.
It is found that $T_{cnt} \approx (5$--$9) \cdot 10^8~{\rm K}$ for neutron 
triplet-state pairing, and $T_{cps} \approx (2$--$3) \cdot 10^9~{\rm K}$ 
for proton singlet-state pairing.
Using observations of 18 isolated neutron stars (including Cas A), 
Ref.~\cite{Beloin:2016zop} found the best-fit values 
$T_{cnt} = 2.09^{+4.37}_{-1.41} \cdot 10^8~{\rm K}$ and
$T_{cps} = 7.59^{+2.48}_{-5.81} \cdot 10^9~{\rm K}$.
On the other hand, the effect of the neutron singlet-state pairing on the temperature
evolution occurs early on, at the neutron star age $\lesssim 100~{\rm yrs}$.
Therefore the Cas A data do not provide useful constraints on $T_{cns}$.

\subsection{Goldstone boson emissivity and the neutron star PBF bound}
\label{sec:GBPBFbound}

\begin{center}
\begin{figure}[t!]
\includegraphics[width=0.6\textwidth,angle=-90]{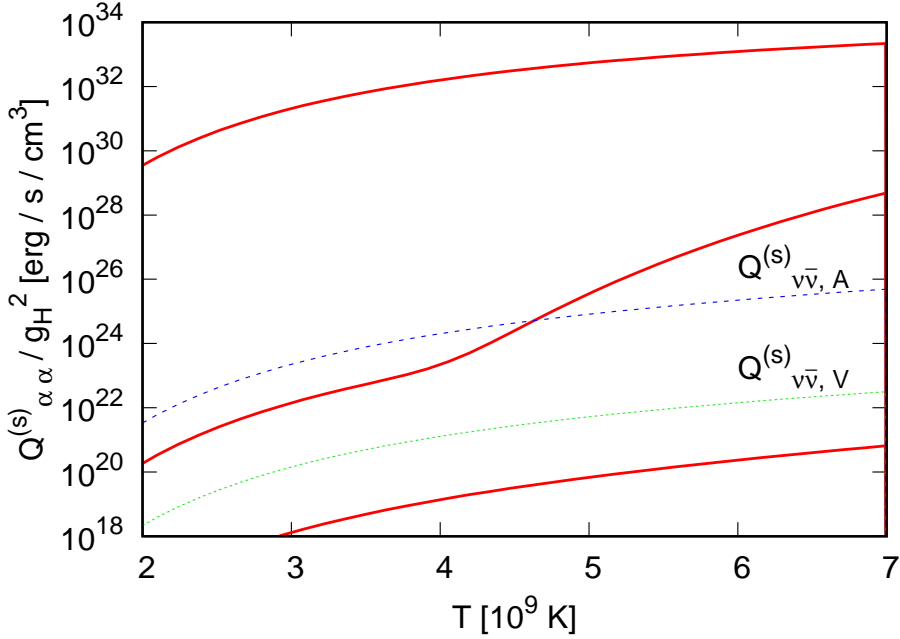}
\caption{Goldstone boson emissivity from the nucleon PBF processes
divided by the Higgs portal coupling $g^2_H$, at different temperatures $T$ below the
neutron singlet-pairing critical temperature $T_{cns}$ (red solid lines).
The gap energy is chosen to be $\Delta_p = 1~\Mev$ at the neutron Fermi momentum
$p_F = 0.77~{\rm fm}^{-1}$, and the critical temperature is approximated by
$T_{cns} \approx 0.5669\, \Delta_p = 7 \cdot 10^9~{\rm K}$.
The nucleon effective mass is fixed at $m^\ast_N = 0.7\, m_N$.
Shown are the results for the three representative cases of the light Higgs boson 
mass and vacuum expectation value (vev) of the radial field $r$: 
$m_h = 1~\Mev$, $\vevr = 100~\Mev$ (top), 
$m_h = 10~\Mev$, $\vevr = 1~\Gev$ (middle), and 
$m_h = 100~\Mev$, $\vevr = 1~\Gev$ (bottom).
Also shown are the vector-current $(Q^{(s)}_{\nu \bar{\nu}, V})$
and the axial-vector current $(Q^{(s)}_{\nu \bar{\nu}, A})$
contributions to the neutrino PBF emissivity in the case of $^1S_0$ pairing.
}
\label{fig:GBemiss_T}
\end{figure}
\end{center}

We numerically evaluate the Goldstone boson emissivity from neutron
singlet-state pairing, $Q^{(s)}_{\al \al}$ in Eq.~(\ref{eq:Q_als}), at radius
$r \simeq 10.2~{\rm km}$ where the neutron Fermi momentum 
$p_F = 0.77~{\rm fm}^{-1}$, and fix the nucleon effective mass at $m^\ast_N = 0.7 m_N$.
We choose a gap energy value of $\Delta_p = 1~\Mev$, which
corresponds roughly to the peak of the AWP3~\cite{Ainsworth:1989yso} 
and WAP~\cite{Wambach:1992ik} gap models considered in 
Ref.~\cite{Ho:2014pta} (cf. Fig.~\ref{fig:Egap_r}.)
For the critical temperature for the neutron singlet-state pairing, 
we adopt the approximate relation given in Ref.~\cite{Ho:2014pta}: 
$T_{cns} \approx 0.5669\, \Delta_p \sim 7 \cdot 10^9~{\rm K}$.
In Fig.~\ref{fig:GBemiss_T} we plot the Goldstone boson emissivity scaled by the 
Higgs portal coupling $g^2_H$ at different temperatures below $T_{cns}$, for three representative cases of the light Higgs boson mass and the vacuum expectation value 
(vev) of the radial field $r$: $(m_h, \vevr) = (1~\Mev, 100~\Mev)$, 
$(10~\Mev, 1~\Gev)$, and $(100~\Mev, 1~\Gev)$. 
Numerically we found in order that the Poisson kernel approximation 
(Eq.~(\ref{eq:PkI1})) works well, the integration over $d E_p$ in 
Eq.~(\ref{eq:Q_als}) should be carried out for $E_p \gtrsim 0.55~m_h$ instead of
$E_p \geq \Delta_p$.
More explicitly, we have 
\begin{equation}
\label{eq:Qals_Pk}
   Q^{(s)}_{\al \al} \approx g^2_H \vevr^2\, F_s\,
   \left(\frac{f_N\, m_N}{m^2_H} \right)^2 \frac{m^\ast_N\, p_F}{4 \pi^3}\, 
   \Delta^2_p\, T \int^\infty_{0.55 m_h / T} \frac{d y\, y}{\sqrt{y^2 - z^2}}
   \frac{1}{(e^y + 1)^2}\, ,
\end{equation}
in the resonance region, where $y \equiv E_p / T$, and $z \equiv \Delta_p / T$.
For $m_h \gtrsim 15~\Mev$, the phase space distribution functions 
$(e^{E_p / T} + 1)^{-2}$ render the resonance enhancement negligible.
Therefore the three solid curves in Fig.~\ref{fig:GBemiss_T} represent three 
distinct cases:
for $m_h = 1~\Mev$, resonance effect can enhance the Goldstone 
boson emissivity in proportion to $\vevr^2 / (1~\Mev)^2$ in the entire temperature 
range shown.
On the $m_h = 10~\Mev$ curve, the transition from the resonance region 
to the large $m_h$ limit (cf. Eq.~(\ref{eq:largemhI1})) is clearly seen around 
$T \sim 4 \cdot 10^9~{\rm K}$.
On the other hand, the temperature in the neutron star superfluid inner crust is too
low for the production of a real light Higgs boson of $m_h = 100~\Mev$, 
so the corresponding curve is completely fixed in the large $m_h$ limit.

Also shown in Fig.~\ref{fig:GBemiss_T} are the vector 
($Q^{(s)}_{\nu \bar{\nu}, V}$, Eq.~(\ref{eq:QnusV})) 
and the axial-vector ($Q^{(s)}_{\nu \bar{\nu}, A}$, Eq.~(\ref{eq:QnusA})) 
contributions to the neutrino PBF emissivity in the case of neutron 
singlet-state pairing. 
We derive constraints on Weinberg's Higgs portal model by finding the model 
parameters $g_H$ and $\vevr$ for each light Higgs boson mass $m_h$, such that
the Goldstone boson emissivity
\begin{equation}
\label{eq:Qcriterion}
   Q^{(s)}_{\al \al} < Q^{(s)}_{\nu \bar{\nu}, V} + Q^{(s)}_{\nu \bar{\nu}, A}\, .
\end{equation}
In the resonance region of producing a real light Higgs boson $h$, where the 
approximation with Poisson kernel limit is applicable, we can simply scale
Eq.~(\ref{eq:Qals_Pk}) by $g^2_H \vevr^2$ to satisfy this criterion.
In Fig.~\ref{fig:NSPBF_gvevr} we present our main result of this work: 
the neutron PBF bounds on $g_H \vevr$ for various $m_h$, determined with \\

{\it i}) $\Delta_p = 1~\Mev$, $p_F = 0.77~{\rm fm}^{-1}$, 
$T_{cns} = 7 \cdot 10^9~{\rm K}$, at $T = 6.85 \cdot 10^9~{\rm K}$; \\
{\it ii}) $\Delta_p = 1.75~\Mev$, $p_F = 0.77~{\rm fm}^{-1}$, 
$T_{cns} = 10^{10}~{\rm K}$, at $T = 6.85 \cdot 10^9~{\rm K}$; \\
{\it iia}) $\Delta_p = 1.75~\Mev$, $p_F = 0.77~{\rm fm}^{-1}$, 
$T_{cns} = 10^{10}~{\rm K}$, at $T = 9.8 \cdot 10^9~{\rm K}$; \\
{\it iii}) $\Delta_p = 1~\Mev$, $p_F = 1.37~{\rm fm}^{-1}$, 
$T_{cns} = 7 \cdot 10^9~{\rm K}$, at $T = 6.85 \cdot 10^9~{\rm K}$. \\

In all cases the nucleon effective mass is fixed at $m^\ast_N = 0.7 m_N$, 
and for simplicity we make the emissivity comparison of at only one temperature 
in each case.
Here case {\it ii}) and {\it iia}) correspond roughly to the peak of the  
CLS~\cite{Cao:2006gq} and the MSH~\cite{Margueron:2007uf} neutron $^1 S_0$ 
pairing gap models considered in Ref.~\cite{Ho:2014pta} (cf. Fig.~\ref{fig:Egap_r}),
at the neutron star radius $r \simeq 10.2~{\rm km}$.
Case {\it iii}) corresponds roughly to the peak of the 
SCLBL~\cite{Schulze:1996zz} gap model at $r \simeq 9.7~{\rm km}$, located
in the neutron star core region.

\subsection{Discussions and comparisons with laboratory and other astrophysics 
bounds}

Comparing among the four cases {\it i}), {\it ii}), {\it iia}), and {\it iii})
listed in the previous subsection, one sees that choosing a higher gap 
energy $\Delta_p$ weakens the PBF bound for $m_h \lesssim 3~\Mev$ by less 
than a factor of $2$.
This is because for $0.55\, m_h < \Delta_p$, the increase of the Goldstone boson emissivity due to larger $\Delta_p$ is less than the increase of the neutrino 
emissivity.
At higher temperatures, the Goldstone boson resonance production is more probable 
due to the larger phase space factor, especially through heavier light Higgs 
boson $h$.
Thus applying the emissivity criterion in Eq.~(\ref{eq:Qcriterion}) at higher
$T$ improves the PBF bound for $m_h \gtrsim 3~\Mev$, and extends  
its validity from $m_h \lesssim 15~\Mev$ to $m_h \sim 20~\Mev$.
Invoking Eq.~(\ref{eq:Qcriterion}) in the neutron star core region 
where the Fermi momentum $p_F$ is larger, the PBF bound can be strengthened 
proportionally for all $m_h$ values.
This is due to the fact that the Goldstone boson emissivity is suppressed more by 
the Fermi velocity than that of the neutrinos: 
$Q^{(s)}_{\al \al} / Q^{(s)}_{\nu \bar{\nu}} \propto v^2_F$.

\begin{center}
\begin{figure}[t!]
\includegraphics[width=0.6\textwidth,angle=-90]{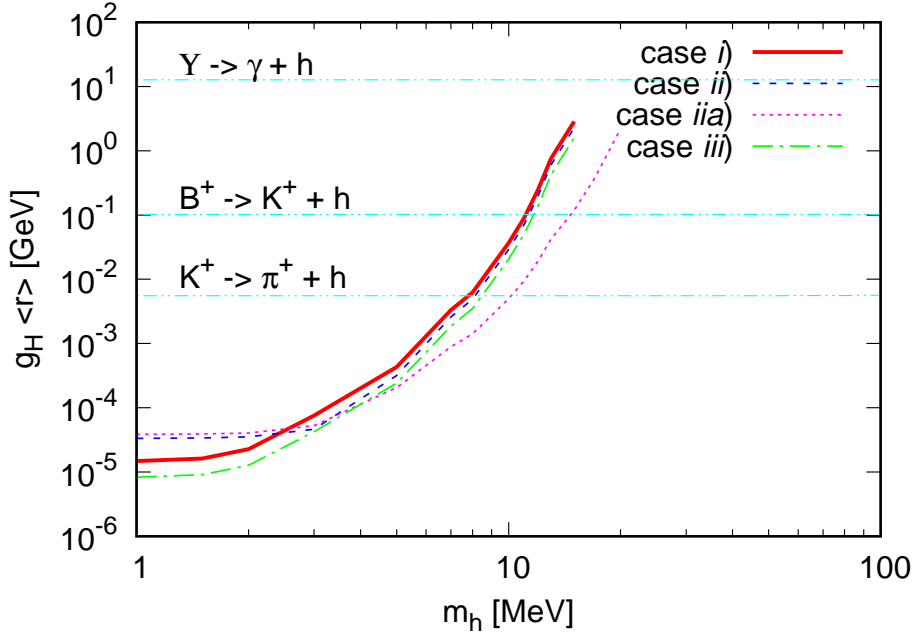}
\caption{Neutron star PBF upper limits on $g_H \vevr$, the product of the Higgs 
portal coupling with the vacuum expectation value of the radial field $r$, 
for various light Higgs boson mass $m_h$.
Several gap models of the neutron singlet-state ($^1 S_0$) pairing 
superfluid are considered at two different radii of the neutron star interior, 
as described in section~\ref{sec:GBPBFbound}.  
Case {\it i}): bound derived by invoking Eq.~(\ref{eq:Qcriterion})
at the neutron star radius $r \simeq 10.2~{\rm km}$ 
where the Fermi momentum is $p_F = 0.77~{\rm fm}^{-1}$, and assuming the
gap energy $\Delta_p = 1~\Mev$, so that the corresponding critical temperature is 
$T_{cns} \approx 7 \cdot 10^9~{\rm K}$. 
The emissivity comparison is made at the crust temperature 
$T = 6.85 \cdot 10^9~{\rm K}$ (red solid line). 
Case {\it ii}): at $p_F = 0.77~{\rm fm}^{-1}$, assuming $\Delta_p = 1.75~\Mev$, 
$T_{cns} \approx 7 \cdot 10^9~{\rm K}$, at $T = 6.85 \cdot 10^9~{\rm K}$ (blue
dashed). 
Case {\it iia}): same as case {\it ii}), but the emissivity comparison is made 
at a higher crust temperature $T = 9.8 \cdot 10^9~{\rm K}$ (pink dotted).
Case {\it iii}): same as case {\it i}), but the comparison is made
at the radius $r \simeq 9.7~{\rm km}$ where $p_F = 1.37~{\rm fm}^{-1}$ 
(green dot-dashed).
Also shown are the upper limits set by laboratory experiments (dash-double dotted
lines, from top to bottom), such as radiative Upsilon decays 
$\Upsilon (n S) \rightarrow \gamma + h$, $B$ meson invisible decay 
$B^+ \rightarrow K^+ + h$, as well as $K$ meson invisible decay
$K^+ \rightarrow \pi^+ + h$.
}
\label{fig:NSPBF_gvevr}
\end{figure}
\end{center}

Also shown in Fig.~\ref{fig:NSPBF_gvevr} are the bounds inferred from the 
experimental results of observing meson invisible decays,
first applied to Weinberg's Higgs portal model 
in Refs.~\cite{Anchordoqui:2013bfa,Huang:2013oua}.
Here the upper limits on the decay branching ratio: 
$\mathcal{B} (B^+ \rightarrow K^+ + h) < 10^{-5}$, 
$\mathcal{B} (K^+ \rightarrow \pi^+ + h) < 10^{-10}$,
as well as
$\mathcal{B} (\Upsilon (n S) \rightarrow \gamma + h) < 3 \cdot 10^{-6}$ are used.
Not shown explicitly in this plot are the collider bound of 
$g_H < 0.011$ in Eq.~(\ref{eq:gcolliderbound}),
and the perturbativity condition in Eq.~(\ref{eq:vevrperturbativitybound}).
The latter can be recast as
\begin{equation}
   g_H \vevr \geq 3.08 \cdot 10^{-6}~{\Gev}\, 
   \left(\frac{g_H}{0.011} \right) \left(\frac{m_h}{1~\Mev} \right)\, ,  
\end{equation}
and one can verify that both of them are satisfied. 
Fig.~\ref{fig:NSPBF_gvevr} demonstrates that the neutron PBF processes in 
neutron stars can indeed probe Weinberg's Higgs portal model in the  
parameter space of low mass ($m_h \lesssim 10~\Mev$), small $\varphi$--$r$ mixing 
angle ($\theta_H \approx 0.0157\, g_H\, (\vevr / 1~\Gev) \lesssim 
10^{-4}~\Gev$), which is unaccessible to current laboratory experiments. 

\begin{center}
\begin{figure}[ht!]
\includegraphics[width=0.6\textwidth,angle=-90]{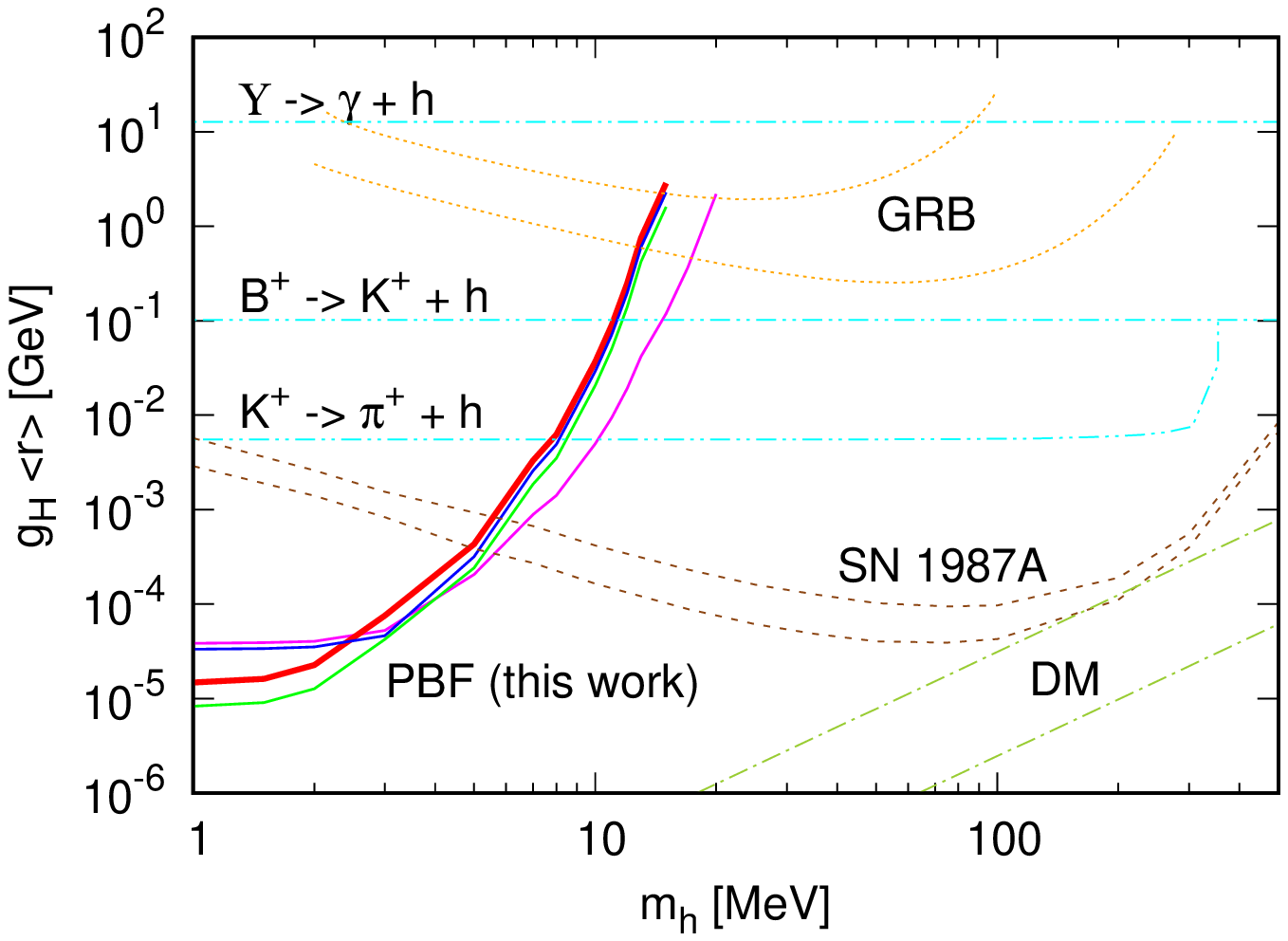}
\caption{Same as Fig.~\ref{fig:NSPBF_gvevr}, here including the upper limits set by
considering other astrophysical objects, as well as by dark matter detection when 
the extended version of Weinberg's Higgs portal model is considered.
Solid lines in red, blue, pink, and green are the neutron PBF bounds derived 
assuming case {\it i}), {\it ii}), {\it iia}), and {\it iii}), respectively.  
The two dashed brown lines below the "SN 1987A" label are those we derived in
Ref.~\cite{Tu:2017dhl} by applying Raffelt's criterion on the energy loss rate of
an exotic particle species in the proto-neutron star core.
Two distinct estimates were adopted for the amplitudes of the nuclear 
bremsstrahlung processes $N N \rightarrow N N \al \al$: 
the global fits for the nucleon-nucleon elastic scattering cross section data (upper), 
and the one-pion exchange (OPE) approximation (lower).
The two dotted orange lines with the "GRB" label between them are those we derived in
Ref.~\cite{Tu:2015lwv} by invoking energy loss argument on the initial fireballs
of gamma-ray bursts. 
Two GRB initial fireball temperature values $T_0 = 18~\Mev$ (lower) and 
$8~\Mev$ (upper) were assumed, and the Higgs portal coupling $g_H$ was taken
to saturate the current collider bound in Eq.~(\ref{eq:gcolliderbound}).
The two dot-dashed yellow-green lines with the "DM" label between them are the upper
limits set by the dark matter direct search experiment LUX, for WIMP mass
$M_\chi = 10~\Gev$ (upper) and $100~\Gev$ (lower), respectively.}
\label{fig:NSGRBSN_gvevr}
\end{figure}
\end{center}

In Fig.~\ref{fig:NSGRBSN_gvevr} we present a comparison of our neutron star 
PBF bounds with other astrophysics bounds we derived previously.
The supernova (SN 1987A) bounds~\cite{Keung:2013mfa,Tu:2017dhl} 
were obtained by invoking Raffelt's analytical criterion on the energy loss rate 
per unit mass due to the emission of an exotic particle species $X$:
$\epsilon_X \lesssim 10^{19}~{\rm erg} \cdot {\rm g}^{-1} \cdot {\rm s}^{-1}$,
at the typical proto-neutron star core temperature $T = 30~\Mev$,
and nuclear density $\rho_0 = 3 \cdot 10^{14}~{\rm g / cm^3}$.
In Ref.~\cite{Tu:2017dhl} we adopted two distinct estimates for the amplitudes 
of the nuclear bremsstrahlung processes $N N \rightarrow N N \al \al$, which are
the one-pion exchange (OPE) approximation, and the global fits for the 
nucleon-nucleon elastic scattering cross section data.
On the other hand, nuclear medium effects were not included.
The gamma-ray burst (GRB) bounds~\cite{Tu:2015lwv} were obtained by invoking the 
energy loss criterion $Q_{e^+ e^- \rightarrow \al \al}\, \Delta t^\prime 
\gtrsim \mathcal{E} / (\Gamma_0\, V_0)$. Here $\Delta t^\prime$ is the time duration
in the fireball comoving frame for the GRB initial fireball to expand from the 
initial radius $R_0$ to $R_0 + \Delta R_0$, $V_0$ is the initial fireball volume,
and $\Gamma_0$ the initial Lorentz factor of the expanding fireball.
The largest uncertainty in this consideration is the unknown initial fireball 
temperature, therefore we assumed two generic values $T_0 = 18~\Mev$ and $8~\Mev$.
 
In the extended version of Weinberg's Higgs portal model, latest exclusion limits
published by the dark matter direct search experiments LUX~\cite{Akerib:2016vxi}, 
PandaX-II~\cite{Tan:2016zwf}, and XENON1T~\cite{Aprile:2017iyp} 
are translated into constraints on the parameter combination 
$f_\chi\, g_H \vevr / m^2_h$ for WIMP mass $M_\chi$ ranging from $6~\Gev$ to
$1~{\rm TeV}$~\cite{Anchordoqui:2013bfa}. 
Shown in Fig.~\ref{fig:NSGRBSN_gvevr} are the DM constraints for $M_\chi = 10~\Gev$ 
and $100~\Gev$, where the WIMP coupling $f_\chi$ is fixed by requiring the 
relic density to be $\Omega_\chi h^2 \simeq 0.11$. 
 
We conclude that in most of the parameter space of Weinberg's Higgs portal model,
the supernova constraints surpass those set by laboratory experiments or by energy
loss arguments in other astrophysical objects, such as the gamma-ray bursts.
Nevertheless, the neutron PBF processes in the superfluid neutron star interior
provide the unique possibility to explore the low mass 
($m_h \simeq \mathcal{O} (1)~\Mev$), small $\varphi$--$r$ mixing angle
($\theta_H \approx g_H\, (\vevr / 1~\Gev) \lesssim 1.6 \cdot 10^{-5}$) region.
DM bounds for $M_\chi \gtrsim 100~\Gev$ remain the strongest constraints 
among all on the extended version of Weinberg's Higgs portal model.

\section{Summary}
\label{sec:summary}

Weinberg's Higgs portal model is another good example to elucidate that high-energy 
astrophysical objects such as the supernovae and gamma-ray bursts are excellent
laboratory for probing particle physics.
In this model, massless Goldstone bosons arising from the spontaneous breaking of 
a $U (1)$ symmetry play the role of the dark radiation in the early universe.
They couple to the Standard Model fields solely through the mixing of the 
$\varphi$ and $r$ fields, which give rise to the SM Higgs boson and a light Higgs
boson $h$. 

Goldstone boson production in the hot proto-neutron star core formed in stellar 
collapse is dominated by the emission of a real light Higgs boson in nuclear bremsstrahlung processes 
and its subsequent decay.
After the neutron star cools to below the critical temperature for the onset of 
superfluidity, resonant production of Goldstone bosons becomes possible again through
the neutron Cooper pair breaking and formation processes.
Theoretical calculation of the neutrino emissivity due to neutron PBF processes 
is a difficult task.
So far all estimates by various approaches agree well, although we note that this 
problem is not settled yet.
In this work we assume that in the superfluid phase, the scalar vertex is 
modified to the same extent by the nuclear medium effects as the vector vertex,
and neglect the unknown shift present in the dressed scalar vertex.

We compare the Goldstone boson emissivity with that of the neutrinos by considering
several superfluid gap models for the neutron singlet-state pairing in the neutron
star inner crust, as well as in the core region.
For a typical gap energy of $1~\Mev$, resonance production of Goldstone boson
pairs is efficient when the light Higgs boson is lighter than about $15~\Mev$,
in which case useful constraints can be obtained.
Assuming a larger gap energy increases the Goldstone boson emissivity, 
but does not improve the PBF constraints.
In gap models which predict higher critical temperatures, the PBF bound can be 
improved and its applicability covers larger light Higgs boson mass $m_h$. 
In those neutron singlet-state pairing gap models which extend to the neutron star
core, the Goldstone boson emissivity is less suppressed due to the larger 
neutron Fermi momentum therein, so the PBF bounds are strengthened.
 
Overall we found that the neutron PBF processes in the superfluid neutron star 
interior offer the unique possibility to explore the low light Higgs boson mass 
($m_h \simeq \mathcal{O} (1)~\Mev$), small $\varphi$--$r$ mixing 
($\theta_H \approx 0.0157\, g_H (\vevr / 1~\Gev) \lesssim 1.6 \cdot 10^{-5}$) 
region in the parameter space of Weinberg's Higgs portal model. 
Together with our previous works on supernovae and gamma-ray bursts constraints,
this study demonstrates further the competitiveness and complementarity of 
astrophysics to laboratory particle physics experiments.

\section*{Acknowledgements}

We thank Dmitry Georgievich Yakovlev and Gordon Baym for the helpful discussions 
and comments.


\begin{thebibliography}{99}

\bibitem{Yakovlev:2000jp}
  D.~G.~Yakovlev, A.~D.~Kaminker, O.~Y.~Gnedin and P.~Haensel,
  Phys.\ Rept.\  {\bf 354} (2001) 1
  [astro-ph/0012122].


\bibitem{Yakovlev:2004iq}
  D.~G.~Yakovlev and C.~J.~Pethick,
  Ann.\ Rev.\ Astron.\ Astrophys.\  {\bf 42} (2004) 169
  [astro-ph/0402143].


\bibitem{Page:2005fq}
  D.~Page, U.~Geppert and F.~Weber,
  Nucl.\ Phys.\ A {\bf 777} (2006) 497
  [astro-ph/0508056].


\bibitem{Potekhin:2015qsa}
  A.~Y.~Potekhin, J.~A.~Pons and D.~Page,
  Space Sci.\ Rev.\  {\bf 191} (2015) no.1-4,  239
  [arXiv:1507.06186 [astro-ph.HE]].


\bibitem{Schmitt:2017efp}
  A.~Schmitt and P.~Shternin,
  arXiv:1711.06520 [astro-ph.HE].



\bibitem{Potekhin:2017ufy}
  A.~Y.~Potekhin and G.~Chabrier,
  Astron.\ Astrophys.\  {\bf 609} (2018) A74
  [arXiv:1711.07662 [astro-ph.HE]].


\bibitem{Ozel:2012wu}
  F.~Ozel,
  Rept.\ Prog.\ Phys.\  {\bf 76} (2013) 016901
  [arXiv:1210.0916 [astro-ph.HE]].



\bibitem{Brown:2017gxd}
  E.~F.~Brown, A.~Cumming, F.~J.~Fattoyev, C.~J.~Horowitz, D.~Page and S.~Reddy,
  Phys.\ Rev.\ Lett.\  {\bf 120} (2018) no.18,  182701
  [arXiv:1801.00041 [astro-ph.HE]].


\bibitem{Ofengeim:2016rkq}
  D.~D.~Ofengeim, M.~Fortin, P.~Haensel, D.~G.~Yakovlev and J.~L.~Zdunik,
  arXiv:1612.04672 [astro-ph.HE].



\bibitem{Ofengeim:2017xxr}
  D.~D.~Ofengeim, M.~Fortin, P.~Haensel, D.~G.~Yakovlev and J.~L.~Zdunik,
  Phys.\ Rev.\ D {\bf 96} (2017) no.4,  043002
  [arXiv:1708.08272 [astro-ph.HE]].


\bibitem{Flowers:1976ux}
  E.~Flowers, M.~Ruderman and P.~Sutherland,
  Astrophys.\ J.\  {\bf 205} (1976) 541.


\bibitem{Voskresensky:1987hm}
  D.~N.~Voskresensky and A.~V.~Senatorov,
  Sov.\ J.\ Nucl.\ Phys.\  {\bf 45} (1987) 411
   [Yad.\ Fiz.\  {\bf 45} (1987) 657].



\bibitem{Yakovlev:1998wr}
  D.~G.~Yakovlev, A.~D.~Kaminker and K.~P.~Levenfish,
  Astron.\ Astrophys.\  {\bf 343} (1999) 650
  [astro-ph/9812366].


\bibitem{Leinson:2006gh}
  L.~B.~Leinson and A.~Perez,
  astro-ph/0606653.


\bibitem{Kolomeitsev:2008mc}
  E.~E.~Kolomeitsev and D.~N.~Voskresensky,
  Phys.\ Rev.\ C {\bf 77} (2008) 065808
  [arXiv:0802.1404 [nucl-th]].


\bibitem{Steiner:2008qz}
  A.~W.~Steiner and S.~Reddy,
  Phys.\ Rev.\ C {\bf 79} (2009) 015802
  [arXiv:0804.0593 [nucl-th]].


\bibitem{Page:2009fu}
  D.~Page, J.~M.~Lattimer, M.~Prakash and A.~W.~Steiner,
  Astrophys.\ J.\  {\bf 707} (2009) 1131
  [arXiv:0906.1621 [astro-ph.SR]].


\bibitem{Kolomeitsev:2010hr}
  E.~E.~Kolomeitsev and D.~N.~Voskresensky,
  Phys.\ Rev.\ C {\bf 81} (2010) 065801
    [arXiv:1003.2741 [nucl-th]].


\bibitem{Kolomeitsev:2010pm}
  E.~E.~Kolomeitsev and D.~N.~Voskresensky,
  Phys.\ Atom.\ Nucl.\  {\bf 74} (2011) 1316
  [arXiv:1012.1273 [nucl-th]].


\bibitem{Leinson:2014cja}
  L.~B.~Leinson,
  Phys.\ Lett.\ B {\bf 741} (2015) 87
  [arXiv:1411.6833 [astro-ph.SR]].


\bibitem{Leinson:2017dlo}
  L.~B.~Leinson,
  arXiv:1712.10214 [nucl-th].


\bibitem{Tamagaki:1970}
  R.~Tamagaki, 
  Prog.\ Theor.\ Phys.\  {\bf 44} (1970) 4.



\bibitem{Shternin:2010qi}
  P.~S.~Shternin, D.~G.~Yakovlev, C.~O.~Heinke, W.~C.~G.~Ho and D.~J.~Patnaude,
  Mon.\ Not.\ Roy.\ Astron.\ Soc.\  {\bf 412} (2011) L108
  [arXiv:1012.0045 [astro-ph.SR]].


\bibitem{Elshamouty:2013nfa}
  K.~G.~Elshamouty, C.~O.~Heinke, G.~R.~Sivakoff, W.~C.~G.~Ho, P.~S.~Shternin, D.~G.~Yakovlev, D.~J.~Patnaude and L.~David,
  Astrophys.\ J.\  {\bf 777} (2013) 22
  [arXiv:1306.3387 [astro-ph.HE]].


\bibitem{Ho:2015vza}
  W.~C.~G.~Ho, C.~M.~Espinoza, D.~Antonopoulou and N.~Andersson,
  Science Adv. 1, e1500578 (2015)
  [arXiv:1510.00395 [astro-ph.SR]].


\bibitem{Ho:2017ipg}
  W.~C.~G.~Ho, C.~M.~Espinoza, D.~Antonopoulou and N.~Andersson,
  JPS Conf.\ Proc.\  {\bf 14} (2017) 010805
  [arXiv:1703.00932 [astro-ph.HE]].


\bibitem{Gezerlis:2014efa}
  A.~Gezerlis, C.~J.~Pethick and A.~Schwenk,
  arXiv:1406.6109 [nucl-th].


\bibitem{Pethick:2015jma}
  C.~J.~Pethick, T.~Schaefer and A.~Schwenk,
  arXiv:1507.05839 [nucl-th].


\bibitem{Haskell:2017lkl}
  B.~Haskell and A.~Sedrakian,
  arXiv:1709.10340 [astro-ph.HE].


\bibitem{Sedrakian:2018ydt}
  A.~Sedrakian and J.~W.~Clark,
  arXiv:1802.00017 [nucl-th].


\bibitem{Keller:2012yr}
  J.~Keller and A.~Sedrakian,
  Nucl.\ Phys.\ A {\bf 897} (2013) 62
  [arXiv:1205.6940 [astro-ph.CO]].


\bibitem{Leinson:2014ioa}
  L.~B.~Leinson,
  JCAP {\bf 1408} (2014) 031
  [arXiv:1405.6873 [hep-ph]].


\bibitem{Sedrakian:2015krq}
  A.~Sedrakian,
  Phys.\ Rev.\ D {\bf 93} (2016) no.6,  065044
  [arXiv:1512.07828 [astro-ph.HE]].


\bibitem{Weinberg:2013kea}
  S.~Weinberg,
  Phys.\ Rev.\ Lett.\  {\bf 110} (2013) no.24,  241301
  [arXiv:1305.1971 [astro-ph.CO]].


\bibitem{Riess:2016jrr}
  A.~G.~Riess {\it et al.},
  Astrophys.\ J.\  {\bf 826} (2016) no.1,  56
  [arXiv:1604.01424 [astro-ph.CO]].


\bibitem{Heavens:2017hkr}
  A.~Heavens, Y.~Fantaye, E.~Sellentin, H.~Eggers, Z.~Hosenie, S.~Kroon and A.~Mootoovaloo,
  Phys.\ Rev.\ Lett.\  {\bf 119} (2017) no.10,  101301
  [arXiv:1704.03467 [astro-ph.CO]].



\bibitem{Keung:2013mfa}
  W.~Y.~Keung, K.~W.~Ng, H.~Tu and T.~C.~Yuan,
  Phys.\ Rev.\ D {\bf 90} (2014) no.7,  075014
  [arXiv:1312.3488 [hep-ph]].


\bibitem{Tu:2017dhl}
  H.~Tu and K.~W.~Ng,
  JHEP {\bf 1707} (2017) 108
  [arXiv:1706.08340 [hep-ph]].


\bibitem{Tu:2015lwv}
  H.~Tu and K.~W.~Ng,
  JCAP {\bf 1603} (2016) no.03,  037
  [arXiv:1512.05165 [hep-ph]].


\bibitem{Sedrakian:2000kc}
  A.~Sedrakian and A.~E.~L.~Dieperink,
  Phys.\ Rev.\ D {\bf 62} (2000) 083002
  doi:10.1103/PhysRevD.62.083002
  [astro-ph/0002228].


\bibitem{Leinson:2000un}
  L.~B.~Leinson,
  Nucl.\ Phys.\ A {\bf 687} (2001) 489
  [hep-ph/0009052].


\bibitem{Schmitt:2005wg}
  A.~Schmitt, I.~A.~Shovkovy and Q.~Wang,
  Phys.\ Rev.\ D {\bf 73} (2006) 034012
  [hep-ph/0510347].


\bibitem{Nambu:1960tm}
  Y.~Nambu,
  Phys.\ Rev.\  {\bf 117} (1960) 648.


\bibitem{Schrieffer:1964zz}
  J.~R.~Schrieffer,
  Front.\ Phys.\  {\bf 20} (2010) 1.


\bibitem{Leinson:2008ba}
  L.~B.~Leinson,
  Phys.\ Rev.\ C {\bf 78} (2008) 015502
  [arXiv:0804.0841 [astro-ph]].


\bibitem{Takatsuka:1992ga}
  T.~Takatsuka and R.~Tamagaki,
  Prog.\ Theor.\ Phys.\ Suppl.\  {\bf 112} (1993) 27.



\bibitem{Levenfish:1994a}
K.~P.~Levenfish and D.~G.~Yakovlev,
Astron.\ Lett.\ 20 (1994) 43.


\bibitem{Levenfish:1994b}
K.~P.~Levenfish and D.~G.~Yakovlev,
Astron.\ Rep.\ 38 (1994) 247.


\bibitem{Sedrakian:2012ha}
  A.~Sedrakian,
  Phys.\ Rev.\ C {\bf 86} (2012) 025803
  [arXiv:1201.1394 [astro-ph.SR]].


\bibitem{Kaminker:1999ez}
  A.~D.~Kaminker, P.~Haensel and D.~G.~Yakovlev,
  Astron.\ Astrophys.\  {\bf 345} (1999) L14
  [astro-ph/9904166].



\bibitem{Leinson:2009mq}
  L.~B.~Leinson,
  Phys.\ Rev.\ C {\bf 79} (2009) 045502
  [arXiv:0904.0320 [astro-ph.HE]].


\bibitem{Sedrakian:2012mv}
  A.~Sedrakian,
  J.\ Phys.\ Conf.\ Ser.\  {\bf 413} (2013) 012024
  [arXiv:1212.0120 [nucl-th]].


\bibitem{Leinson:2009nu}
  L.~B.~Leinson,
  Phys.\ Rev.\ C {\bf 81} (2010) 025501
  [arXiv:0912.2164 [astro-ph.SR]].


\bibitem{Leinson:2010pk}
  L.~B.~Leinson,
  Phys.\ Rev.\ C {\bf 82} (2010) 065503
   Erratum: [Phys.\ Rev.\ C {\bf 84} (2011) 049901]
  [arXiv:1012.5387 [hep-ph]].



\bibitem{Leinson:2011jr}
  L.~B.~Leinson,
  Phys.\ Rev.\ C {\bf 84} (2011) 045501
  [arXiv:1110.2145 [nucl-th]].



\bibitem{Leinson:2016dat}
  L.~B.~Leinson,
  arXiv:1611.03794 [nucl-th].



\bibitem{Cheung:2013oya}
  K.~Cheung, W.~Y.~Keung and T.~C.~Yuan,
  Phys.\ Rev.\ D {\bf 89} (2014) no.1,  015007
  [arXiv:1308.4235 [hep-ph]].


\bibitem{Anchordoqui:2013pta}
  L.~A.~Anchordoqui and B.~J.~Vlcek,
  Phys.\ Rev.\ D {\bf 88} (2013) 043513
  [arXiv:1305.4625 [hep-ph]].



\bibitem{Anchordoqui:2013bfa}
  L.~A.~Anchordoqui, P.~B.~Denton, H.~Goldberg, T.~C.~Paul, L.~H.~M.~Da Silva, B.~J.~Vlcek and T.~J.~Weiler,
  Phys.\ Rev.\ D {\bf 89} (2014) no.8,  083513
  [arXiv:1312.2547 [hep-ph]].



\bibitem{Serot:1994xh}
  B.~D.~Serot and H.~B.~Tang,
  Phys.\ Rev.\ C {\bf 51} (1995) 969
  [nucl-th/9404023].



\bibitem{Saito:2003js}
  K.~Saito and K.~Tsushima,
  Phys.\ Lett.\ B {\bf 575} (2003) 4
  [nucl-th/0307053].


\bibitem{Kaminker:2001ag}
  A.~D.~Kaminker, P.~Haensel and D.~G.~Yakovlev,
  Astron.\ Astrophys.\  {\bf 373} (2001) L17
  [astro-ph/0105047].


\bibitem{Kaminker:2001eu}
  A.~D.~Kaminker, D.~G.~Yakovlev and O.~Y.~Gnedin,
  Astron.\ Astrophys.\  {\bf 383} (2002) 1076
  [astro-ph/0111429].


\bibitem{Andersson:2004aa}
  N.~Andersson, G.~L.~Comer and K.~Glampedakis,
  Nucl.\ Phys.\ A {\bf 763} (2005) 212
  [astro-ph/0411748].


\bibitem{Ho:2014pta}
  W.~C.~G.~Ho, K.~G.~Elshamouty, C.~O.~Heinke and A.~Y.~Potekhin,
  Phys.\ Rev.\ C {\bf 91} (2015) no.1,  015806
  [arXiv:1412.7759 [astro-ph.HE]].


\bibitem{Oertel:2016bki}
  M.~Oertel, M.~Hempel, T.~Klähn and S.~Typel,
  Rev.\ Mod.\ Phys.\  {\bf 89} (2017) no.1,  015007
  [arXiv:1610.03361 [astro-ph.HE]].


\bibitem{Lorene}
LORENE (Langage Objet pour la RElativit$\acute{e}$ Num$\acute{e}$riquE),
https://lorene.obspm.fr/


\bibitem{Hempel:2009mc}
  M.~Hempel and J.~Schaffner-Bielich,
  Nucl.\ Phys.\ A {\bf 837} (2010) 210
  [arXiv:0911.4073 [nucl-th]].


\bibitem{Typel:2009sy}
  S.~Typel, G.~Ropke, T.~Klahn, D.~Blaschke and H.~H.~Wolter,
  Phys.\ Rev.\ C {\bf 81} (2010) 015803
  [arXiv:0908.2344 [nucl-th]].



\bibitem{Fischer:2013eka}
  T.~Fischer, M.~Hempel, I.~Sagert, Y.~Suwa and J.~Schaffner-Bielich,
  Eur.\ Phys.\ J.\ A {\bf 50} (2014) 46
  [arXiv:1307.6190 [astro-ph.HE]].


\bibitem{Steiner:2007rr}
  A.~W.~Steiner,
  Phys.\ Rev.\ C {\bf 77} (2008) 035805
  [arXiv:0711.1812 [nucl-th]].


\bibitem{Page:2010aw}
  D.~Page, M.~Prakash, J.~M.~Lattimer and A.~W.~Steiner,
  Phys.\ Rev.\ Lett.\  {\bf 106} (2011) 081101
  [arXiv:1011.6142 [astro-ph.HE]].


\bibitem{Beloin:2016zop}
  S.~Beloin, S.~Han, A.~W.~Steiner and D.~Page,
  arXiv:1612.04289 [nucl-th].



\bibitem{Ainsworth:1989yso}
  T.~L.~Ainsworth, J.~Wambach and D.~Pines,
  Phys.\ Lett.\ B {\bf 222} (1989) 173.


\bibitem{Wambach:1992ik}
  J.~Wambach, T.~L.~Ainsworth and D.~Pines,
  Nucl.\ Phys.\ A {\bf 555} (1993) 128.
  
  
\bibitem{Cao:2006gq}
  L.~G.~Cao, U.~Lombardo and P.~Schuck,
  Phys.\ Rev.\ C {\bf 74} (2006) 064301
  [nucl-th/0608005].  


\bibitem{Margueron:2007uf}
  J.~Margueron, H.~Sagawa and K.~Hagino,
  Phys.\ Rev.\ C {\bf 77} (2008) 054309
  [arXiv:0712.3644 [nucl-th]].


\bibitem{Schulze:1996zz}
  H.~J.~Schulze, J.~Cugnon, A.~Lejeune, M.~Baldo and U.~Lombardo,
  Phys.\ Lett.\ B {\bf 375} (1996) 1.


\bibitem{Huang:2013oua}
  F.~P.~Huang, C.~S.~Li, D.~Y.~Shao and J.~Wang,
  Eur.\ Phys.\ J.\ C {\bf 74} (2014) no.8,  2990
  [arXiv:1307.7458 [hep-ph]].


\bibitem{Akerib:2016vxi}
  D.~S.~Akerib {\it et al.} [LUX Collaboration],
  Phys.\ Rev.\ Lett.\  {\bf 118} (2017) no.2,  021303
  [arXiv:1608.07648 [astro-ph.CO]].


\bibitem{Tan:2016zwf}
  A.~Tan {\it et al.} [PandaX-II Collaboration],
  Phys.\ Rev.\ Lett.\  {\bf 117} (2016) no.12,  121303
  [arXiv:1607.07400 [hep-ex]].


\bibitem{Aprile:2017iyp}
  E.~Aprile {\it et al.} [XENON Collaboration],
  Phys.\ Rev.\ Lett.\  {\bf 119} (2017) no.18,  181301
  [arXiv:1705.06655 [astro-ph.CO]].


\end{thebibliography}

\end{document}